# Rapid Differentiation between Microplastic Particles Using Integrated Microwave Cytometry with 3D Electrodes

Yagmur Ceren Alatas[1,2], Uzay Tefek[1,2], Sayedus Salehin[1,2], Hashim Alhmoud[1,2] and M. Selim Hanay[1,2,*]

1 Department of Mechanical Engineering, Bilkent University, 06800 Ankara, TURKEY
2 UNAM – Institute of Materials Science and Nanotechnology, Bilkent University, 06800, Ankara TURKEY

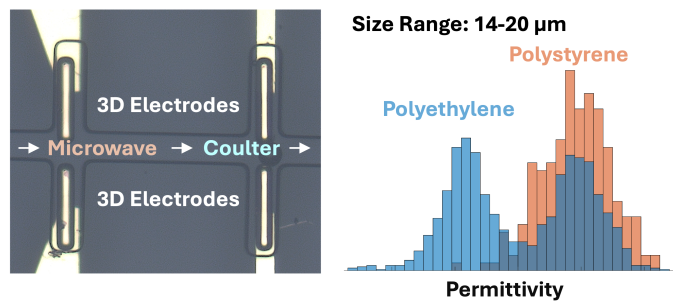

**ABSTRACT:** Rapid identification of microparticles in liquid is an important problem in environmental and biomedical applications such as for microplastic detection in water sources and physiological fluids. Existing spectroscopic techniques are usually slow and not compatible with flow-through systems. Here we analyze single microparticles in the 14 – 20 micrometer range using a combination of two electronic sensors in the same microfluidic system: a microwave capacitive sensor and a resistive pulse sensor. Together, this integrated sensor system yields the effective electrical permittivity of the analyte particles. To simplify data analysis, 3D electrode arrangements were used instead of planar electrodes, so that the generated signal is unaffected by the height of the particle in the microfluidic channel. With this platform, we were able to distinguish between polystyrene (PS) and polyethylene (PE) microparticles. We showcase the sensitivity and speed of this technique and discuss the implications for the future application of microwave cytometry technology in the environmental and biomedical fields.

The identification of microparticles in a given sample is an important problem for screening of consumer products, drugs, and environmental samples. Indeed, for environmental screening, the presence of microplastics in water samples has increasingly been seen as an emerging, critical problem. Approaches for microplastic identification currently involve time-consuming and challenging operations [1]. For microplastic identification, two techniques are used predominantly: Fourier Transform Infrared Spectroscopy (FTIR) and microRaman Spectroscopy [2]. The typical size resolution of FTIR is 20 μm which can be improved by the use of special setups that increase the cost and duration for sample preparation. Micro-Raman Spectroscopy is similarly a time-consuming technique (estimated to take 10 minutes per particle analysis on average [1]). While microRaman Spectroscopy can reach down to 1 μm in resolution, it reportedly does not work well for particles with dark colors [2].

Apart from the spectroscopy techniques, pyrolysis gas chromatography mass spectrometry [3] provides chemical information about the constituents of a given sample and supports microplastic quantification efforts. However, this technique does not provide size distribution information which is critical since the transportation and toxicity properties of microparticles strongly depend on their sizes.

Recently, electronic detection techniques have been considered for the identification of microplastics [4-6]. By using conventional impedance cytometry at two frequencies (10 kHz and 1.1 MHz), a recent work [4] demonstrated the separation of microplastics from biological species such as seeds and organisms in the 300 μm – 1 mm range. While this study played a pivotal role in the introduction of microplastic processing workflows with electronic sensors, it explored a relatively large range of particle sizes (300 μm – 1 mm). However, most of the microplastic particles in potable water for example, appear to be in the range of 20 μm and smaller [7], e.g. due to removal of larger particles by filtering. It should also be noted that the differentiation between plastics and biological species is facilitated by the large water content (and hence unusually high permittivity) of biological species. The real challenge, however, is to identify microplastics from microparticles with similar dielectric permittivity values, as was recently shown in a study differentiating between microglass and microplastic particles [5].

One limitation of classical impedance cytometry is the frequency of operation. Normally at DC and RF frequencies, the ions in the liquid sample contribute to the electronic response by generating a current under the electrical field. This ionic motion generally obfuscates the capacitive response of the analytes which carries the permittivity information needed for microparticle identification. Thus, the identification capability provided by utilizing the dielectric permittivity of microparticles as a metric is missed by sensors operating at low frequencies. To overcome this problem, sensors operating at microwave frequencies have been developed. Early work in the microfluidics-integrated microwave sensors focused on detecting single cells, micro and nanoparticles [8-25]. More recently however, microwave sensors used multiple sensor modalities to obtain multi-dimensional information such as the location sensing of droplets [26], microscopy-supported classification for material content [27], and machine-learning assisted identification of cellular nuclei by microwave sensors [28].

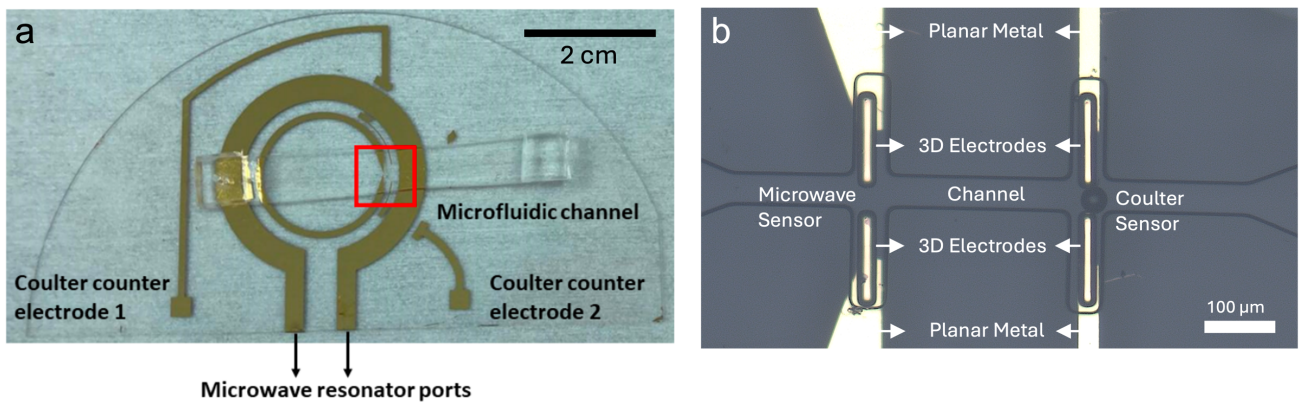

**Figure 1.** The integrated sensing platform with microwave and Coulter sensors. (a) Overall picture of the chip indicating the two rings forming the resonator and electrodes for the coulter. (b) Zoomed-in view of the sensing region where the gray looking metal-coated polymer structures are connected on the 2D metals. The microfluidic channel passes through the sensing regions of both sensors.

To measure the dielectric constants of microparticles, a new sensor family was developed [5] where a microwave sensor is integrated with a low-frequency resistive pulse sensor (i.e. a Coulter counter). With this architecture, the low-frequency sensor detects the volume of a particle; simultaneously, the microwave sensor detects the product of the geometric volume and the Clausius-Mossotti factor of the particle (which is a function of the permittivity of the particle.) This way, dividing the microwave signal by the Coulter signal obtained from the two sensors yields the Clausius-Mossotti factor of the particles, which is then used to differentiate different particles from each other. This platform was also used to study the effects of chemicals in modifying the internal structure of single cells as well [5] which was not possible previously.

One challenge with the abovementioned microwave sensors is the two-dimensional nature of typical microwave resonators. While 2D resonators are easy to fabricate (e.g. by standard photolithography), the electric field generated has an inhomogeneous 3D distribution inside a microfluidic channel: as a result, same particle flowing at different altitudes will generate signals of different magnitudes. To overcome this limitation, and decrease the burden on data analysis for height compensation [29, 30], we developed microwave sensors with 3D electrodes surrounding the sensing region. Initially, a liquid metal (Galinstan) was used to fill dead-end microfluidic channels to establish 3D electrodes [31] and characterize 30 μm particles. Later, an elastomer structure (made out of the polymer SU8) in a microfluidic system was metal coated to generate 3D electrodes [32] and distinguish polystyrene particles of different size (12 μm vs 20 μm) without the need for extensive data analysis. Both systems were microwave-only sensors, and they only yielded the microwave size of microparticles as a result. However, the integration of microwave sensors with resistive pulse sensors both having 3D electrodes remained an important development to tackle. By using both electronic sensors simultaneously as in [5], and introducing 3D electrodes, it now becomes possible to obtain material-specific information from a sensor directly with minimal data analysis requirements.

In this work, we demonstrated the integration of microwave and Coulter sensors with 3D electrode sensing regions in a microfluidic system. We first discuss the fabrication of the platform, and then provide details about the electronic detection. We quantify the performance of the sensor on a polystyrene (PS) sample first consisting of particles in the 14-20 μm range. Next, we analyzed a mixture of polystyrene (PS) and polyethylene (PE) microparticles (10-20 μm range). The mixture analysis indicated a clear separation of PE and PS particles. The relative dielectric constant of PE and PS are close to each other (bulk values are 2.3 and 3.0, respectively): therefore, the differentiation between microplastic particles of two different compositions at the single-particle level signifies an important improvement for practical usage of this platform.

**Sensor Design**

The sensor architecture is a combination of a split ring resonator (SRR) microwave sensor and a resistive pulse sensor operating at several hundred kHz (Fig. 1). The active regions of both sensors are encapsulated by the same microfluidic channel. The active regions are designed to be close to each other (300 μm apart), so that a particle passing through the system gets detected by both sensors sequentially and in quick order. This design enables events occurring in both sensors to be matched together. However, the inter-sensor distance was not decreased below 300 μm to avoid cross-coupling between the two sensors. Most of the sensor structure is formed by thin gold layers (120 nm), however at the sensing region, both sensors have extra 3D features with rectangular prismatic shapes. These features are structures (SU8) coated with metal to act as 3D electrodes. This way, the electric field across the electrode pairs are more or less uniform, and the signals due to particles generate height-independent signals.

The microwave resonator is composed of two concentric rings [14, 33]. The outer ring is connected to the outside instrumentation by an SMA connector and is mainly used to excite the resonance in the inner ring via inductive coupling. The internal ring acts as an RLC resonator, where the inductance originates from the currents flowing through the ring, and the capacitance is predominantly determined by the gap in the ring. The microfluidic channel is aligned so that the fluid flow passes through this gap in the ring, thereby sampling the electric field where it has the largest electrical field intensity.

The electrodes of the low-frequency sensor extend between the concentric rings due to geometric constraints. These electrodes are then wire-bonded to auxiliary metal pads that transmit the signal to the edge of the chip where they can be connected to the external instrumentation. One connection is used to provide a voltage at the Coulter electrode, whereas the other connection is used to measure the current.

**Sensor Fabrication**

The fabrication procedure of the sensor is similar to the method described earlier [32]. A 4-inch, 500 µm thick fused silica wafer was diced into two equal halves using a dicing saw. The first step involved patterning coplanar SRR and Coulter counter electrodes using UV photolithography. The pattern was metalized using a thermal evaporator by depositing 120 nm Gold on a 10 nm Chromium adhesion layer. Coplanar electrodes were obtained after acetone lift-off.

SU8 pillars, which serve as the structural material for the 3D electrodes, were patterned at the tip of the SRR and Coulter counter electrodes, forming the sensing region. SU8-2050 was spin-coated onto the substrate (Step 1: 500rpm, 100 acceleration, 20 seconds; Step 2: 3500 rpm, 300 acceleration, 1 minute) and exposed to UV light after soft baking using a pre-patterned electrode mask. The SU8 microelectrodes were designed as rectangular prismatic structures with a height of approximately 45 µm. After development and hard baking, a positive photoresist layer was patterned under the same mask used for SU8 electrodes to obtain a lift-off layer after SU8 electrode metallization. AZ5214E photoresist was spin-coated using 3000 rpm speed, 300 rpm acceleration, and for 40 seconds. The final thickness of the photoresist was around 1.4 µm at this stage — less than the usual thickness used for patterning the coplanar electrodes previously so that the resist can spread easily between the 3D electrodes at the sensing region. Before the SU8 electrode metallization step, oxygen plasma cleaning was performed to remove any dirt or dust that may cause difficulty for lift-off.

The last step of sensor fabrication was the metallization of the SU8 pillars. The most critical aspect of this step was to obtain a conformal deposition profile on the SU8 pillars to create a uniform electric field inside the microfluidic channel. Although sputter coating yielded a conformal deposition profile, it was not suitable for lift-off because the deposition profile did not leave any area for acetone to enter and remove the resist in the post-metallization step. When using sputter coating, metal remained in certain areas of the resonator surface, causing circuit shortages.

As an alternative, SU8 microelectrodes were coated with gold through thermal evaporation [34]. Gold does not adhere to fused silica without an adhesion layer, which makes the lift-off process easier still. However, since the coating profile obtained by thermal evaporation is anisotropic compared to sputter coating, the substrates were tilted 45° on the thermal evaporation holder so that a conformal deposition profile could be obtained on the inner and outer walls of the SU8 electrodes (Fig. 2). Deposition was repeated twice to coat the walls of the SU8 electrodes on opposite sides. After the first deposition, the chamber was vented, and the tilt direction of the substrates was inverted before the second coating step (Fig. 2b).

The coating rate was adjusted to 0.4 Å/s during deposition to prevent excessive heating and charring of the underlying SU8. During the process, the chamber pressure was set to $5 \times 10^{-5}$ Torr, and the final gold thickness for each deposition step was 40-45 nm (two of those deposition steps were conducted for each sample at different angles to cover all the faces of the SU8 3D structures). Finally, the substrates were immersed in acetone for metal lift-off for a duration of 2 hours, and subsequently checked under a microscope.

Once the electronic sensors with 3D electrodes were patterned, a PDMS microfluidic channel was attached. Microfluidic channel mold fabrication was carried out using SU8 2050 on a Silicon wafer with a final thickness of 50 μm. The spin speed at the second step was 3400 rpm, less than the spin speed used for the SU8 electrode step earlier, so that the microfluidic channel height is greater than the 3D electrodes and no air gap remains between the SU8 electrodes and the channel upon bonding. SU8 was exposed to UV light under the microfluidic channel mask after soft baking and placed on a hot plate for post-exposure baking. Finally, after hard baking, PDMS was poured and allowed to cure. The PDMS channel was peeled off from the mold upon cross-linking. The PDMS channel was then bonded to the resonator substrate with alignment using oxygen plasma treatment. Oxygen plasma treatment was done using 350W power, 40 seconds, 40 sccm $O_2$ and 10 sccm $N_2$ flow. The microfluidic channel was aligned with the SU8 microelectrodes manually under an optical microscope.

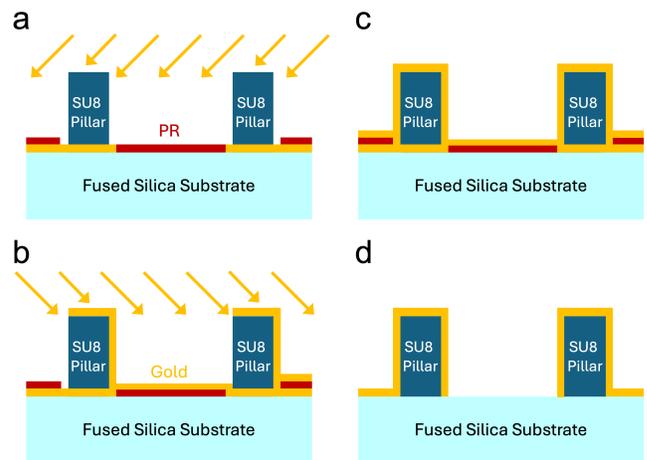

**Figure 2.** Fabrication of the 3D electrodes at the sensing region. (a) After the SU8 pillars are deposited on the coplanar electrodes, the sample is coated with photoresist (PR) and tilted by 45° before gold metal evaporation. (b) After the first deposition, the sample is tilted to the opposite direction by 45° and gold deposition is repeated. (c) After the two depositions, internal faces of both SU8 pillars are coated. (d) Lift-off removes the PR and the excess gold around the device to yield 3D electrodes which are connected to planar gold at the bottom.

**Electronic Characterization**

In testing the system, we first characterized the performance of the two sensors separately. For the SRR microwave resonator, the resonance frequency and quality factor of different modes were measured by using a vector network analyzer (VNA). The spectrum hosted a collection of modes with relatively high quality factors (Fig. 3a).

Since the resonance frequency of the SRR was above the operational range of the lock-in amplifiers, an external heterodyne RF circuit was employed [5, 8-10, 25, 35] as shown in Fig. 3b. This circuit tracks both the amplitude and phase of a microwave resonator as a response to a constant drive frequency close to its resonance. The drive is provided by the signal generator. Part of the drive is reserved for mix-down detection at the end, while the other part is modulated by a low-frequency lock-in reference and is provided to the microwave resonator through a circulator. The response of the resonator is read back at the reflection through the circulator and modulated down to the RF frequencies of the lock-in amplifier using an IQ – mixer. The down-mixed components are provided to the lock-in amplifier to measure the in-phase and quadrature components of the sensor response (typically with a 1 ms time constant). These components are used to calculate the imaginary part of the sensor response in the presence of the microwave background [5, 32]. The data sampling rate was set to 13.4 kSa/s.

For the Coulter sensor, a frequency of operation at 360 kHz was used. In this case, a lock-in amplifier was used to directly provide the excitation voltage on one of the electrodes, and the resulting current was read through a transimpedance analyzer on the other electrode (Fig. 3b). During the particle sensing experiments, the lock-in time constant for the Coulter was also set to 1 ms, while the gain of the transimpedance amplifier was set to 10 kΩ, and the data transfer rate was 14.4 kSa/s. Similar to the microwave resonator, the drive frequency was kept fixed, and the passage of the microparticles was observed by the rapid spike-like events that were generated. With this setup, microparticles of different sizes (in the range of 14 – 20 μm) were passed through the sensing region.

Microwave sensor and Coulter signals were simultaneously analyzed with a custom-written MATLAB script to obtain the size and dielectric permittivity information of a microparticle species. The initial step in the MATLAB script involved resampling microwave and Coulter signals at 10 kSa/s because the sampling rates of the Coulter and microwave signals were different during the experiments. Data acquisition for microwave and Coulter signals were started manually during the experiments, and this added a small time delay between the signals. Indices of the same event in Coulter and microwave spectra were matched to synchronize the signals. Baseline reduction was applied to the signals to compensate for the effect of long-term drift. Coulter amplitude signal was normalized by dividing the waveform by the baseline value.

The microwave amplitude was used to determine particle-induced shifts using the built-in *findpeaks* MATLAB function. The microwave phase signal was utilized to identify potentially valid events. Matching the microwave phase and amplitude peaks was conveniently done, since they occur almost simultaneously. However, matching Microwave and

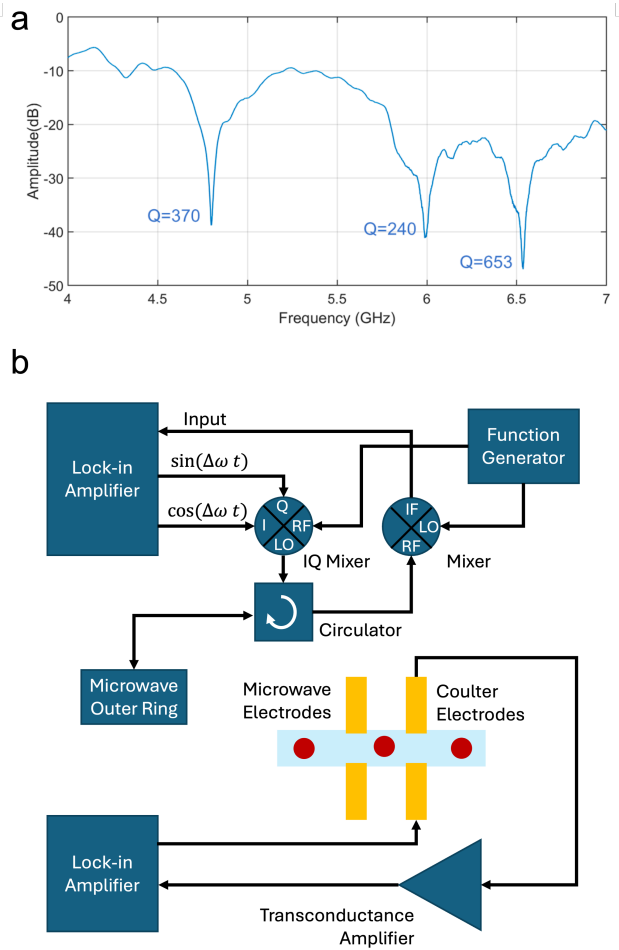

**Figure 3.** The measurement of the sensor. (a) Open loop response of the microwave resonator. Quality factors of different modes were calculated by the 3-dB method and listed on the graph. (b) The measurement circuitry combining both microwave and Coulter sensing. On the microwave side, the signal is fed and read from the outer ring of the SRR, which is inductively coupled to the inner ring where the microwave electrodes are located.

Coulter signals was more challenging since variations in flow rate during the experiment affected the delay between the microwave and Coulter signals. To solve this problem, a 200 ms window around each candidate event was checked in the Coulter signal to locate a single matching peak for the same event. If no corresponding peak or multiple peaks were detected within this window, the candidate event was considered invalid and removed from further analysis.

The raw data signals obtained in both Coulter and microwave sensor measurements are proportional to the volume of the particles. For ease of comparison, we took the cube root of the peak magnitudes: the cube root of the Coulter signal corresponds to the geometric diameter and the cube root of the imaginary part of the frequency shift corresponds to the electric diameter. For measurements with different particles, a microparticle with known size and dielectric permittivity was used for calibration.

The ratio of electrical diameter to geometrical diameter is referred to as the opacity of a particle. Considering that the imaginary part of the microwave shift is proportional to the normalized capacitance change, the ratio of opacity for different microparticles is expected to be proportional to the cube root of the ratio of their Clausius-Mossotti factor. For an experiment involving two different microparticle species, the opacity histogram is therefore expected to show a bimodal distribution.

**Characterization of Microplastic Mixture**

We used the platform to analyze two microplastic polymers closely related to each other. In this case, we chose two microplastics with similar dielectric constants: 3.0 (PS, PSMS-1.07) and 2.3 (PE, WPMS -1.00 PE, Cospheric). We used the platform to conduct two measurements in sequence: 1) a PS only measurement, 2) a mixture of PS and PE measurements. By performing measurements on the mixture, we can directly observe if the two types can be differentiated from each other. By using the PS-only measurements, we can identify the space for the parameter region of PS particles. The two experiments are shown in the Coulter (x-axis) vs. microwave (y-axis) plane in Fig. 4a. The PS only run is shown in orange: in this run about 100 PS microparticles were measured ranging in size between 14 and 20 μm. The mixture experiment is shown in blue: in this case the PS sample was mixed with the PE sample that had a size range of 10-20 μm. In this case, two separate sub-populations appeared in the 2D plot (Fig. 4a). One of the sub-populations matches well with the PS-only measurements, while the other sub-population appears above this PS population. This second population is attributed to the PE particles, since PE particles have lower permittivity and as a result generate a larger contrast (i.e. larger electrical diameter) in electrical measurements with a water medium which has a high permittivity (~78). Moreover, the PE sample covers a wider size range (10-20 μm) compared to the PS sample (14-20 μm), so the extension of the PE sample towards the left side of the x-axis is well explained by its relative polydispersity (x-axis).

Based on the ratio between the microwave and Coulter signals, the opacity values are extracted and plotted in Fig. 4b. Here, the PS-only experiment (orange) produces a single peak; whereas in the mixture experiment (blue), the emergence of the two sub-populations is again evident. The results demonstrate that the sensing platform can differentiate between PS and PE microplastic particles.

Considering that the mean value for PS occurs near 1 a.u., the peak of the PE corresponds to 1.2 a.u. Typically, in analyzing electrical measurements, the Maxwell-Garnett mixing formula, as well as the Clausius-Mossotti Factors of the materials, are used to describe the relationship between the permittivity of the material and the sensor output (such as opacity) when the volume fraction of the analyte particle inside the medium is very small. However, in our case the volume fraction of the particle reaches to 5% of the sensing region which violates the assumptions of Maxwell-Garnett model [36]. Indeed, the observed ratio between the opacity values of the two particles is well explained by a parallel-plate capacitor model owing to the geometry defined by 3D electrodes. The inclusion of a particle inside a capacitor results in the *series addition* of the dielectric constants, which

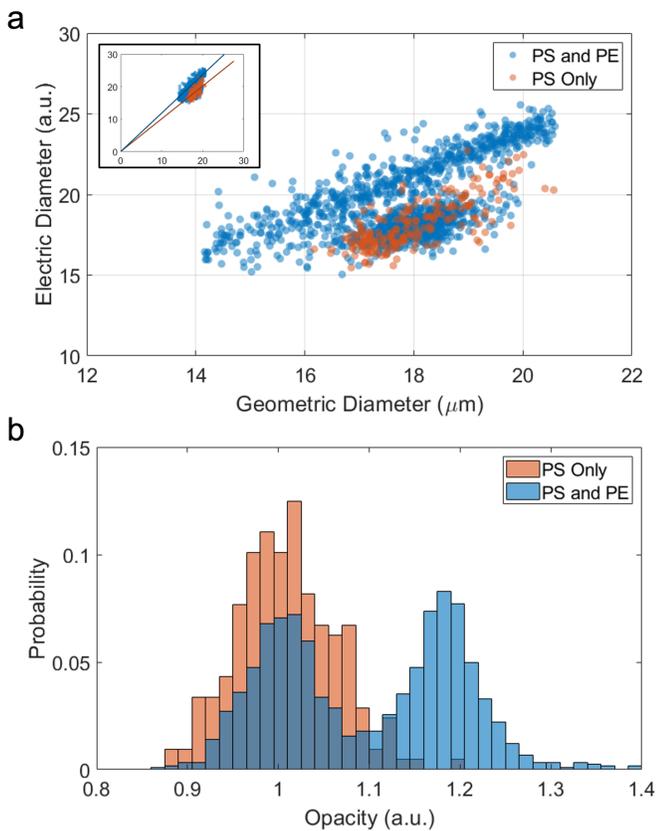

**Figure 4.** Electronic detection of microplastics mixture. (a) Scatter plot for PS only (orange) and PS-PE mixture (blue) samples. In the mixture data, there are two subpopulations visible: one subpopulation overlaps with the PS-only region. The other subpopulation is higher in terms of electrical diameter (i.e. lower permittivity) – as expected of PE particles. The inset shows the data zoomed out to reveal the location of the scatter plot with respect to the origin. (b) Opacity values for the PS-only (orange) and the PS-PE mixture (blue). The separation of PS and PE particles is clear in the mixture data where one peak matches with the PS-only experiment, and a second peak emerges which correspond to the opacity values of PE particles.

then changes the effective capacitance:

$$C_{total} \propto \left(\frac{V_{particle}}{\epsilon_{particle}} + \frac{V_{medium}}{\epsilon_{medium}}\right)$$

In this equation, $\epsilon_{particle}$ and $\epsilon_{medium}$ are the permittivity values of the particle and medium, respectively; $V_{particle}$ is the particle volume, $V_{medium}$ is the volume remaining between the electrodes after subtracting the particle volume. The signal measured by the microwave sensor is the change in the total capacitance as a new particle enters into the sensing volume. By comparing the ratio of capacitance changes for two particles of the same size but different permittivity levels, we obtain an opacity ratio of 1.14 by using the nominal bulk values of PS and PE. The expected ratio (1.14) is close to the observed value of in the experiments (1.18). This observation, together with the clear separation between PS and PE at the single-particle level demonstrated the utility of this approach for material characterization.

**Conclusion**

In this work, we describe a new sensing platform combining a high-frequency microwave resonator and a low-frequency resistive pulse sensor operating within a small microfluidic sensing region for microparticle permittivity characterization. The sensing region comprises 3D electrode geometries that ensure the creation of uniform electric fields throughout the microfluidic sensing region. This echoes earlier work with 2D electrodes [5] where the electrical permittivity of the analyte microparticles was measured directly through the combination of low and high frequency sensors. Due to the planar electrode geometry in [5], the non-uniform electric field necessitated positional compensation for the analyte particles in the sensing region which in turn increased analysis complexity and reduced sensitivity. The 3D electrode geometry in this work greatly reduced the analysis complexity and time, and even allowed for the differentiation of two polymeric materials (PS and PE) with very similar permittivity values. This is the first time as far as we are aware, where two very similar materials (PS and PE) were characterized and differentiated in an entirely electronic manner without relying on analytical chemistry methods. Future improvements in the fabrication of 3D electrodes with embedded microfluidic channels will further increase the sensitivity and throughput of the current sensing platform and allow rapid and real-time data analysis. We anticipate that this would lead to the development of small form-factor electronic sensors for environmental and biomedical *in situ* applications.

**AUTHOR INFORMATION**

**Corresponding Author**

M. Selim Hanay: selimhanay@bilkent.edu.tr

**Funding Sources**

This project has received funding from the European Research Council (ERC) under the European Union's Horizon Europe programme by Proof-of-Concept Grant RAMP-UP Grant No: 101113438.